\documentclass[12pt]{article}
\usepackage{CJK}
\usepackage{amsthm,amsmath,bm,amsfonts}
\usepackage{multirow,graphicx,subfigure,lscape}
\usepackage{}
\usepackage[colorlinks]{hyperref}

\newenvironment{changemargin}[2]{%
\begin{list}{}{%
\setlength{\topsep}{0pt}%
\setlength{\leftmargin}{#1}%
\setlength{\rightmargin}{#2}%
\setlength{\listparindent}{\parindent}%
\setlength{\itemindent}{\parindent}%
\setlength{\parsep}{\parskip}%
}%
\item[]}{\end{list}}

\def\x{{\sf x}}
\def\y{{\sf y}}
\def\v{{\sf v}}
\def\w{{\sf w}}

\def\I{{\sf \xi}}
\def\hat{\widehat}

\def\xb{{\bf   x}}
\def\Var{{\rm var}}

\def\all{{\rm all}}
\def\sel{{\rm sel}}
\def\resid{{\rm resid}}
\def\kernel{{\rm kernel}}

\newtheorem{thm}{Theorem}
\newtheorem{cor}{Corollary}

\newcommand{\goes}{\rightarrow}
\newcommand{\nn}{\nonumber}
\newcommand{\ve}{\varepsilon}
\newcommand{\K}{\kappa}
\newcommand{\X}{\x}

\newcommand{\xmat}{X}

\def\cA{{\cal A}}
\def\cI{{\cal I}}

\newcommand{\alg}{Alg }
\title{Random Partitioning and Distribution-based Thresholding for Iterative Variable Screening\\ in High Dimensions}
\author{Yu-Hsiang Cheng$^{\text{a}}$ \hspace{1cm}Tzee-Ming Huang$^{\text{b}}$\thanks{Corresponding author. E-mail address: tmhuang@nccu.edu.tw.} \hspace{1cm}Su-Yun Huang$^{\text{a}}$\\ \\
$^{\text{a}}$Institute of Statistical Science, Academia Sinica, Taiwan  \\ $^{\text{b}}$Department of Statistics, National Chengchi University, Taiwan}
\date{}
\begin{document}
\maketitle
\begin{abstract}
In big data analysis, a simple task such as linear regression can become very challenging as the variable dimension $p$ grows. As a result, variable screening is inevitable in many scientific studies. In recent years, randomized algorithms have become a new trend and are playing an increasingly important role for large scale data analysis. In this article, we combine the ideas of variable screening and random partitioning to propose a new iterative variable screening method. For moderate sized $p$ of order $O(n^{2-\delta})$, we propose a basic algorithm that adopts a distribution-based thresholding rule. For very large $p$, we further propose a two-stage procedure. This two-stage procedure first performs a random partitioning to divide predictors into subsets of manageable size of order $O(n^{2-\delta})$ for variable screening, where $\delta >0$ can be an arbitrarily small positive number. Random partitioning is repeated a few times. Next, the final estimate of variable subset is obtained by integrating results obtained from multiple random partitions. Simulation studies show that our method works well and outperforms some renowned competitors. Real data applications are presented. Our algorithms are able to handle predictors in the size of millions. 
\end{abstract}

\section{Introduction}
In recent years and for high dimensional data analysis, the problem of variable screening and selection has been receiving increasing attention, especially in the analysis of large scale data sets such as gene sets (e.g.~\cite{he2011},~\cite{liu2015},~\cite{gan2017}).   When the number of variables is very large,  it is common to adopt a procedure that includes multiple stages of variable screening and selection. E.g.,  Wasserman and Roeder~\cite{was2009} propose a three-stage procedure consisting of  variable screening in the first two stages and followed by variable cleaning in the third stage. In this article,  we focus on the variable screening problem  for the linear regression model and propose a two-stage method. Consider
\begin{equation} \label{eq:model}
\y = \sum_{j=1}^p \beta_j \x_j +\ve =\beta^\top \xb +\ve,
\end{equation}
where $\y$ is  the response variable, $\xb=(\x_1,\dots,\x_p)^\top$ is the vector of $p$ predictors, $\beta=(\beta_1, \ldots, \beta_p)^\top$ is the vector of coefficients and $\ve$ is the error, which has mean zero. For simplicity but without loss of generality, we assume the above linear regression model has zero intercept.  Here $p$ is much larger than $n$, but only a few predictors are influential to $\y$ (see, e.g.,~\cite{tib1994} and~\cite{zou2006}).  

For variable screening, a well-known screening method is sure independence screening (SIS), proposed by Fan and Lv~\cite{fan2008}, where predictors
with large  absolute sample correlations to the response variable are selected. In the same article, Fan and Lv  also propose an extension of SIS, called  iterative sure independence screening (ISIS), to improve  selection accuracy.  
At each iteration of ISIS,  first SIS is performed to reduce the number of predictors,  and next as a second step,  variable selection via  LASSO or SCAD is applied to the predictors obtained from SIS. Their screening is based on the absolute sample correlation between a predictor and the residuals obtained  from regressing the response on the predictors selected at the previous iteration. The iterative procedure stops when a pre-specified number of predictors is reached.
   
While ISIS performs much better than SIS according to the simulation results in~\cite{fan2008}, its implementation requires one to determine the number of predictors to be selected and the tuning parameter(s) in the second step of variable selection. It requires extra efforts in computation as well as in hyper-parameter selection. For instance, if one would like to use ISIS with LASSO and would select the $\lambda$ parameter in LASSO by cross-validation at each iteration, then it can be computationally quite expensive. To solve the difficulty encountered in ISIS, we make the following changes and propose a new variable screening method in Section~\ref{sec:method}. Our method combines the ideas of variable screening and randomized algorithms for large scale data analysis~\cite{lo2009,halko2011,woodruff2014}.

\begin{enumerate}
\item 
We propose new thresholding rule (Subsection~\ref{sec:in_control}). This rule adopts a distribution-based threshold for variable screening. 

\item We develop two new algorithms for moderate-sized $p$ and large $p$, respectively.  We briefly describe the two algorithms below. 
\begin{itemize}
\item 
{\sf Moderate-size condition}. We say that $p$ satisfies the moderate-size condition,
if $p \leq Cn^{2-\delta}$ for some constant $C>0$ and some small $\delta >0$.

{\sf Basic Algorithm} (Subsection~\ref{sec:iter}). Based on the new thresholding rule, the Basic Algorithm performs iterative variable screening for moderate-sized $p$.
\item
{\sf Large-size condition}. We say that $p$ satisfies the large-size condition,
if $p/n^{2-\delta}\to\infty$ for any $\delta >0$.

{\sf Two-stage Algorithm} (Subsection~\ref{sec:large_p}). For large $p$, predictors are randomly partitioned into subgroups of moderate size, so that the Basic Algorithm can be applied. The random partitioning is repeated a few times and results from these multiple random partitions are integrated to form the final estimate of active predictor set. 
\end{itemize}
\end{enumerate}

The rest of the article is organized as follows.  In Section~\ref{sec:method}, we propose a new method for variable screening and give two theorems for theoretical justification. We provide two  screening algorithms: one is the {Basic Algorithm} for moderate-sized $p$ and the other is the main {Two-stage Algorithm} for large $p$. Simulation results are given in Section~\ref{sec:sim}.  Two real data applications are given in Section~\ref{sec:real_data}. Concluding remarks are given in Section~\ref{sec:conclusion}.  Proofs are given in Section~\ref{sec:proof}.

\section{Method} \label{sec:method}
\subsection{Distribution-based thresholding} \label{sec:in_control}
Consider the screening problem for the regression model~(\ref{eq:model}). Let $\cA$ and $\cI$ denote, respectively, the active set and inactive set of predictors for regression model~(\ref{eq:model}),
\begin{equation}
\cA=\left\{\x_j:\beta_j \neq 0\right\}~~{\rm and}~~ 
\cI=\left\{\x_j:\beta_j = 0\right\}.
\end{equation} 
Suppose that we have $n$ IID observations, $\{( \x_{i1}, \ldots, \x_{ip},\y_i)\}_{i=1}^n$, from model~(\ref{eq:model}). Let $Y_n = (\y_1, \ldots, \y_n)^\top$ and let $\hat{F}_j$ be the marginal empirical CDF of $\{\x_{ij}\}_{i=1}^n$ for $j=1,\dots,p$. 
A predictor $\x_j$ is included into the active set estimate if the absolute sample correlation between $(\x_{1j}, \ldots, \x_{nj})^\top$ and $Y_n$ exceeds the $(1-\alpha)$ quantile of the conditional distribution of $G(\cdot | Y_n, \hat{F}_1, \ldots, \hat{F}_p)$, which is given by
\begin{eqnarray}
&& G\left(t \big|Y_n, \hat{F}_1, \ldots, \hat{F}_p\right),~~\forall t \in{\mathbb R}   \nonumber \\
&& =P\left\{ \left.  \max_{1 \leq j \leq p} \left| {\rm corr}(Y_n,\X^*_j )\right| \le t
\right| Y_n, \{( \x_{i1}, \ldots, \x_{ip})\}_{i=1}^n 
\right\}   \label{eq:thresh1}  
 \\ && =P\left\{ \left.  \max_{1 \leq j \leq p} \left| {\rm corr}(Y_n,\X^*_j )\right| \le t \right| Y_n, \hat{F}_1, \ldots, \hat{F}_p \right\} . \nonumber
\end{eqnarray}
Here auxiliary variables $\X^*_1,\ldots,\X^*_p$ are conditionally independent of $Y_n$ given predictor observations $\{( \x_{i1}, \ldots, \x_{ip})\}_{i=1}^n$, and  each 
  $\X^*_j=(\x_{1j}^*, \ldots, \x_{nj}^*)^\top$ is a random sample of size $n$ from $\hat{F}_j$ given  $\{( \x_{i1}, \ldots, \x_{ip})\}_{i=1}^n$.  The screening threshold, the $(1-\alpha)$ quantile, is denoted by
\begin{equation}\label{eq:quantile}
t_{\alpha,Y_n, \hat{F}_1, \ldots, \hat{F}_p} = G^{-1}(1-\alpha|Y_n, \hat{F}_1, \ldots, \hat{F}_p).
\end{equation}
Note that $t_{\alpha,Y_n, \hat{F}_1, \ldots, \hat{F}_p}$ is a random variable depending on the empirical data. To save notation, we will use $t_{\alpha,n}$ for simplicity.
When $n$ is small, $t_{\alpha,n}$ can be calculated using bootstrap approximation. When $n$ is large, normal approximation can be used. Given $Y_n$, $\hat{F}_1$, $\ldots$ and $\hat{F}_p$, the conditional distribution of the sample correlation between $Y_n$ and $\X^*_j$ is asymptotically $N(0, 1/n)$ for large $n$.  Motivated by the asymptotic normality result, we also consider replacing the screening threshold $t_{\alpha,n}$ by 
\[
 z(n, p, \alpha) = \Phi^{-1} \left( 1-0.5(1-   (1-\alpha)^{1/p}  ) \right) \frac{1}{\sqrt{n}},
\]
where $\Phi$ is the CDF of standard normal. Note that we will refer to $z(n, p, \alpha)$ as the normal approximation screening threshold because it is suggested by normal approximation, but we do not claim that $z(n, p, \alpha)$ can approximate $t_{\alpha,n}$ well in any sense. We simply use $z(n, p, \alpha)$ as another screening threshold.

Below we state a theorem and its corollary, which together ensure that an important predictor will be selected with probability tending to one when using $t_{\alpha,n}$ or $z(n, p , \alpha)$ as the screening threshold and $p$ satisfying the moderate-size condition.

\begin{thm} \label{thm:1}
Aassume that
\[
 \max_{1\leq j \leq p}~ \frac{1}{n} \sum_{i=1}^n \left( \frac{\x_{ij} - \mu_{j,n}}{ \sigma_{j,n}} \right)^6  = O_p(1),
\]
where  $\mu_{j,n} = \sum_{i=1}^n \x_{ij}/n$ and $\sigma_{j,n}^2= \sum_{i=1}^n (\x_{ij}-\mu_{j,n})^2/n$.
Also assume that $p$ satisfies the moderate-size condition. Then we have, for $t >0$,
\[P(t_{\alpha,n} < t) \goes 1~~\mbox{for every $t>0$}.\]
\end{thm}

Let $ {\hat \cA}_n$  denote the estimated active set, where the screening threshold can be $t_{\alpha,n}$ or $z(n,p,\alpha)$.  We have the following corollary.
\begin{cor} \label{cor:1} Assume that the active set size is fixed and finite, i.e., $|\cA|=\kappa<\infty$. Then, we have
\begin{equation}
\lim_{n\to\infty} P\left\{\cA \subset   {\hat \cA}_n   \right\} =1.
\end{equation}
\end{cor}
The proofs of Theorem~\ref{thm:1} and Corollary~\ref{cor:1} are given in Section~\ref{sec:proof}.

While Theorem~\ref{thm:1} and Corollary~\ref{cor:1} ensure that, with probability tending to one, no important variables are missed, it is also of interest to know  whether it is likely to include  too many unimportant predictors using $t_{\alpha,n}$ or $z(n, p, \alpha)$  as the screening threshold.  In the following theorem, we give bounds on  the
probability of selecting at least $r$ unimportant predictors  when $z(n, p, \alpha)$ is used as the screening threshold.  
 It is assumed in the following theorem that  $\x_1$, $\ldots$, $\x_p$ and $\y$ are normally distributed.  
\begin{thm} \label{fa:consist}
Suppose that  among the $p$ predictors $\x_1$, $\ldots$, $\x_p$, $p-\K$ predictors are independent of $\y$. Assume that $\x_1$, $\ldots$, $\x_p$ and $\y$ are normally distributed.  Let $N$ be the number of unimportant predictors that are selected, then the distribution of $N$ is $Bin(p-\K,p_1)$ with  success probability $p_1$ given by
\begin{equation} \label{eq:fap1}
 p_1 = P \left( \left| \frac{Z}{ \sqrt{Z^2 + U_n} } \right| > \frac{c(p)}{\sqrt{n}} \right),
\end{equation}
where $Z$ and $U_n$ are independent variables with distributions $N(0,1)$ and $\chi^2(n-2)$ respectively,
and where
\begin{equation} \label{eq:facp}
 c(p) = \Phi^{-1}\left( 1- 0.5\left[1-(1-\alpha)^{1/p}\right]\right).
\end{equation}
Assume that $\kappa$ is fixed, $\lim_{n\goes \infty} p = \infty$ and $p=O(n^{2-\delta})$ for some arbitrarily small $\delta >0$. Then, there exists $p_1^* \geq p_1$ such that
$\lim_{n \goes \infty} pp_1^* = -\ln(1-\alpha) \stackrel{\rm def}{=}\lambda_0$. 
Let  ${\hat\cA}_n(z(n, p, \alpha))$ denote the estimated active set with screening threshold $z(n, p, \alpha)$. 
Then for large enough $n$ and $p_1^* \in [0,1]$, we have
\[
P\left\{ \left| {\hat\cA}_n(z(n, p, \alpha))\cap \cI \right|\ge r\right\}
\le 1 - \sum_{\ell=0}^{r-1} \frac{(p-\K)!}{(p-\K-\ell)!\ell !} (p_1^*)^{\ell} (1-p_1^*)^{p-\K-\ell},
\]
which tends to 
\[
  1 - \sum_{\ell=0}^{r-1} \frac{e^{-\lambda_0}\lambda_0^{\ell}}{ \ell !}  \leq \frac{\lambda_0^r}{r!}
\]
as $n\goes \infty$. 
\end{thm}
The proof of Theorem~\ref{fa:consist} is given in Section~\ref{sec:proof}.

\noindent {\bf Remark}. The idea of using auxiliary predictor variables independent of the response to set a thresholding rule is not new. See, e.g.,  the sure independent ranking and screening (SIRS) by Zhu et al.~\cite{zhu2011}. A predictor ranking score is used to measure the contribution of each variable, and a random screening threshold is used.  The random threshold is the maximum of the ranking scores of $d$ auxiliary variables.  Our usage of auxiliary variables $\x_1^*,\dots,\x_p^*$ in (\ref{eq:thresh1}) plays a similar role as the auxiliary variables in SIRS. However, our thresholding rule is different from SIRS.

\subsection{Iterative screening for moderate $p$} \label{sec:iter}
As Theorem~\ref{thm:1} and Theorem~\ref{fa:consist} are valid only for $p$ of order $O(n^{2-\delta})$, our first algorithm is designed for such $p$. The proposed algorithm is iterative and based on the threshold described in Subsection~\ref{sec:in_control}.
When the predictors are correlated, it is possible that some important predictors are only weakly correlated with 
the response, but they may have stronger association with the response jointly. In such a case, performing iterative screening can help improve the detection power. This point is mentioned in both~\cite{fan2008} and~\cite{zhu2011}. 

To develop our algorithms, below we first define two functions {\bf DB-SIS} and  {\bf Resid}, which stand for distribution-based SIS and taking residuals. Suppose that $Y$ is an $n$-vector,  $S\subset \{ \x_1, \ldots \x_p \}$ is a predictor subset and $\alpha\in(0, 1)$. 
\begin{itemize}
\item   
  {\bf DB-SIS}($Y$, $S$, $\alpha$) is the set $\{\x_j\in S: |\rho_j| >  G^{-1}(1-\alpha|Y, \hat{F}_{\x_{j_1}}, \ldots, \hat{F}_{\x_{j_k}})\}$, where $\rho_j$ is the correlation between $Y$ and $\x_j$ and $S=\{\x_{j_1},\dots, \x_{j_k}\}$. 
\item  
{\bf Resid}($Y$, $S$) is the vector of residuals by regressing $Y$ on variables in $S$ .
\end{itemize}
Our first algorithm is stated as follows.  For convenience, we set the constant $C$, which appears in the moderate-size condition, to one.

\noindent\rule{\textwidth}{1pt}
{\bf Basic Algorithm (for $\bm{p \leq n^{2-\delta}}$).} Suppose that $\x_1$, $\ldots$, $\x_p$ are predictors  and $Y_n$ is the observation vector of the response variable. Let $S_{\all}$ denote the collection of all $p$ predictors. $\alpha \in (0, 1)$ is pre-specified.
\begin{enumerate}\itemsep=0pt
\renewcommand{\labelenumi}{(\theenumi)}
\item Let $S_{\sel} =$ DB-SIS$(Y_n, S_{\all}, \alpha)$. 

\item  \label{st:2} Let $S_{\sel}^c = S_{\all} \setminus S_{\sel}$ be the set of predictors that are not yet selected. Let $Y_{\resid} = $Resid$(Y_n, S_{\sel})$. 

\item \label{st:3} Add the output of DB-SIS$(Y_{\resid}, S_{\sel}^c)$ to $S_{\sel}$.   

\item   Carry out Steps (\ref{st:2}) and (\ref{st:3}) until no more important predictors can be found or  the $Y_{\resid}$ in Step (\ref{st:2}) is a vector of zeros. 

\item  The predictors selected  are the predictors in $S_{\sel}$. Output $\hat\cA_n =S_{\sel}$.
\end{enumerate}
\rule{\textwidth}{1pt}

\begin{itemize}
\item Note. As mentioned in previous section,  $G^{-1}(1-\alpha|Y_n, \hat{F}_1, \ldots, \hat{F}_p)$ can be approximated using bootstrap approximation when $n$ is small, or can be approximated using the normal approximation $z(n,p,\alpha)$ when $n$ is large. In {\bf DB-SIS}, the approximation of $G^{-1}(1-\alpha|Y, \hat{F}_{\x_{j_1}}, \ldots, \hat{F}_{\x_{j_k}})\}$ can be obtained similarly. In our simulation studies, we use normal approximation for $n \geq 200$ and bootstrap approximation for $n < 200$.
\end{itemize}

\subsection{Screening with random partitioning for large $p$} \label{sec:large_p}
In this subsection, we describe our second screening algorithm, which is designed for the large $p$ case.  When $p$ is very large, the screening threshold $G^{-1}(1-\alpha|Y_n, \hat{F}_1, \ldots, \hat{F}_p)$ can be improper, as Theorem~\ref{thm:1} and Theorem~\ref{fa:consist} are no longer valid. In such a  case, we can partition the predictors into smaller subgroups and perform screening on subgroups. Indeed, $\{\x_1, \ldots, \x_p\}$  will be randomly partitioned into $k$ disjoint subsets of sizes $p_1$, $\ldots$, $p_k$, where $p_1$, $\ldots$, $p_k$ are approximately equal size and each $p_i$ does not exceed $n^{2-\delta}$. Here $\delta$ is pre-specified  and we take $\delta= 0.03$ in all of our simulation studies.  The random partitioning will be repeated for $T$ times.  An additional second stage   integration of results is included to form the final estimate of active predictor set.  Below is our second screening algorithm. Here, for large $p$ case, we assume $n\geq 200$ and use the normal approximation for setting the screening threshold.

\noindent\rule{\textwidth}{1pt}
{\bf Two-stage Algorithm  (for large $\bm{p}$).} Suppose that $\x_1$, $\ldots$, $\x_p$ are predictors and $Y_n$ denotes the observation vector of the response variable. Suppose $p>n^{2-\delta}$.
\begin{enumerate}\itemsep=0pt
\renewcommand{\labelenumi}{(\theenumi)}
\item  {\sc First stage screening.}  Let $\Omega =  \{\x_1, \ldots, \x_p\}$. For $t=1$, $\ldots$, $T$, carry out the tasks in  (a)-(d) below, and get $\hat\cA_{n,t}$.
\begin{enumerate}\itemsep=0pt
\item Partition $\Omega$ into $k$ disjoint  subsets $\Pi_1$, $\ldots$, $\Pi_k$. Take $S_{\kernel} = \emptyset$, $S_{\sel} =\emptyset$  and $Y_{\resid}=Y_n$.
\item For $\nu=1$, $\ldots$, $k$, let $\Pi_\nu^* = \Pi_\nu \setminus S_{\kernel}$ and carry out the tasks in i and ii below:
\begin{enumerate}\itemsep=0pt
\item Let $A_\nu =$ DB-SIS$(Y_{\resid}, \Pi_\nu^*, \alpha)$. Add $A_\nu$ to $S_{\sel}$.
\item Calculus the adjusted $R^2$ when regressing $Y_n$ on predictors in $A_\nu \cup S_{\kernel}$. Denote the corresponding adjusted $R^2$ by $R^2_{adj, \nu}$.
\end{enumerate}
\item Let $A^*$ be the $A_\nu$ with the largest $R^2_{adj, \nu}$ in Step (b) among $A_1$, $\ldots$, $A_k$. Add the predictors in $A^*$ to $S_{\kernel}$.
  \item Take $Y_{\resid} =$Resid$(Y_n, S_{\kernel})$.
  \item Repeat Steps (b)-(d) until no more predictors are added to $S_{\sel}$, or the largest adjusted $R^2$ does not increase, or the number of predictors in $S_{\sel}$ exceeds $n$.
  \item   For each random partition  we have $\hat\cA_{n,t} \stackrel{\rm def}{=} S_{\sel}$. 
\end{enumerate}
\item  {\sc Second stage  integration.}  For $\ell= 2$, $\ldots$, $T$, let $\Psi_\ell$ be the collection of the predictors that appear exactly $\ell$ times in these $T$ sets $\hat\cA_{n,1}, \ldots, \hat\cA_{n,T}$. Take $\Psi^* = \Psi_T$  and $Y_{\resid} =$Resid$(Y_n, \Psi^*)$.
\begin{enumerate}\itemsep=0pt
\item For $\ell = T-1$, $\ldots$, 2 (in descending order), if $\Psi_\ell$ is nonempty, carry out the tasks in i and ii below: 
\begin{enumerate} \itemsep=0pt
\item Regress $Y_{\resid}$ on each predictor in $\Psi_\ell$ and add the predictor to $\Psi^*$ if the corresponding $p$-value is less than 0.05. 
\item Take $Y_{\resid} =$Resid$(Y_n, \Psi^*)$.
\end{enumerate}
\item The final estimate of active predictor set $\hat\cA_n=\Psi^*$.
\end{enumerate}
\end{enumerate}
\vspace{-0.6cm}\rule{\textwidth}{1pt}

\section{Simulation studies} \label{sec:sim}
\subsection{Experimental settings} \label{sec:dgp}
In this section, we carry out some simulation studies with different settings to check the performance of the proposed screening procedures in Subsections~\ref{sec:iter} and ~\ref{sec:large_p}.  Throughout Section~\ref{sec:sim} (except for Subsection~\ref{sec:heavy}), the $p$ predictors $\x_1, \ldots, \x_p$ are generated from $N(0,1)$ with specific covariance matrix $\Sigma$, and for each predictor, $n$ observations are generated. 
Here we consider three kinds of $\Sigma$.
\begin{eqnarray*}
\Sigma_1 &=&I, \qquad
\Sigma_2 =\Big[0.75^{|i-j|}\Big]_{i,j=1}^p,\\[1.5ex]
\Sigma_3 &=&  \left\{
       \begin{array}{ll}
       \rho_1, & \mbox{for } i \leq 10,~ j \leq 10,\\
       0.05, & \mbox{for } 11 \leq i \leq p,~ 11 \leq j \leq p,\\
       0, & \mbox{otherwise.}
       \end{array} \right. 
\end{eqnarray*}
The same types of covariance matrices are also considered in~\cite{zhu2011}. The observations of the response $\y$ is generated according to (\ref{eq:model}), 
where $\beta^\top = (\beta_1, \ldots, \beta_{10}, 0, \ldots, 0)$ is the coefficient vector and the $\beta_i$s are generated from the  uniform distribution on $[0.5, 1.5]$. The errors are generated independently from $N(0, \sigma^2)$, where the variance $\sigma^2$ is determined via the following equation with a pre-specified $r^*$:
\[ \frac{E\big(\Var(\beta^\top \xb|\beta )\big)}{E\big(\Var(\beta^\top \xb |\beta )\big)+\sigma^2} = r^*. \] 
Note that when $\beta$ is not random, $r^*$ is the $R^2$ for the linear model~(\ref{eq:model}), which is used in~\cite{zhu2011} to indicate the relative noise level. In order to evaluate the selection accuracy for a given variable selection method result $\hat{\cA}$, we use the following accuracy measure
\begin{equation} \label{acc_measure}
\frac{|\cA \cap \hat{\cA}|}{|\cA|},
\end{equation}
where $\hat{\cA}$ is the active set estimate, i.e., the collection of the variables selected by the selection method, and $\cA$ is the true active set. In our simulation studies, $\cA = \{\x_1, \ldots, \x_{10}\}$ and $|\cA| = 10$. 
We compare the performance of our screening method with those of
LASSO, adaptive LASSO, and ISIS-SCAD. For LASSO and adaptive LASSO, the tuning parameter $\lambda$ of the $L_1$ penalty can be determined by cross-validation. However,   according to our empirical experience,  using cross-validation to choose $\lambda$ for LASSO often leads to including too many predictors, when the predictors are independent. Thus in our simulation studies, the penalty parameter $\lambda$ for LASSO and adaptive LASSO is chosen in such a way that the number of predictors selected is approximately the same as (or a little bit larger than)  the number of predictors selected by our method, so that the results for LASSO and adaptive LASSO can be compared to those for our method on a fair basis.
For ISIS-SCAD, we use the function \verb+SIS+ in the R package \verb+SIS+.  All the default settings are used and the method for tuning the regularization parameter is set to SCAD.   

\subsection{The case of moderate $p$}  \label{sec:small_p_sim}
In this subsection,  we present numerical results for the case $p\le n^{2-\delta}$   with $\delta= 0.03$. We apply Basic Algorithm to four cases of combinations of $n$ and $p$: $(n, p) \in \{ (100, 8700)$, $(200, 34000)$, $(300, 75000)$, $(400, 133000)\}$, where $p$ is close to $n^{1.97}$. 
When the sample size $n$ is relative small ($n < 200$), we use bootstrapping approximate for the screening threshold~(\ref{eq:quantile}). When $n\ge 200$,  we use normal approximation $z(n, p, \alpha)$. The value of $\alpha$ is set to $0.5$. We carry out 500 replicate runs for each combination of $(n, p)$. 
The data are generated according to the settings in Subsection~\ref{sec:dgp}.
The $\rho_1$ in $\Sigma_3$ is 0.5 or 0.3. 
When $\Sigma=\Sigma_1$,  we consider $r^* = 0.91$ or 0.95. When $\Sigma=\Sigma_2$ or $\Sigma=\Sigma_3$, all four  methods perform very well with $r^*>0.9$ and it is difficult to compare the four methods with $r^*>0.9$. Thus we consider $r^* = 0.5$ or 0.55 for $\Sigma=\Sigma_2$ and $\Sigma=\Sigma_3$.

The simulation results are given in Table~\ref{simu1}. The average  accuracy~(\ref{acc_measure}) obtained over  500 runs is reported for four methods: our Basic Algorithm, LASSO, adaptive LASSO and  ISIS-SCAD.  The median number of predictors selected by each method is also included in parentheses. Our Basic Algorithm outperforms the other three in almost all cases, especially when the sample size $n$ is small. In the case where $n=100$ and $\Sigma=\Sigma_3$ with $\rho_1 = 0.3$, ISIS-SCAD performs slightly better than our Basic Algorithm. The performance of all  four methods improves as $r^*$ or $n$ increases.

\begin{landscape}
\begin{table}[ht] \renewcommand{\arraystretch}{1.1} \small
\tabcolsep=5.5pt
\begin{center}
    \begin{tabular}{cccccccccccc}
    \hline  \hline
$\Sigma$ & $r^*$ &  & $n=100$ & $n=200$ & $n=300$ & $n=400$ & $r^*$ & $n=100$ & $n=200$ & $n=300$ & $n=400$ \\
\cline{4-7}
\cline{9-12}
\multirow{4}{*}{$I_{p \times p}$} & \multirow{4}{*}{0.91}  & Basic \alg & 0.6112(9) & 0.9984(12) & 1(12) & 1(12) &\multirow{4}{*}{0.95}& 0.7768(12) & 1(12) & 1(11) & 1(12)\\   
& & LASSO & 0.5198(10) & 0.8902(12) & 0.9886(12) & 0.9990(12) && 0.5926(12) & 0.9476(12) & 0.9988(11) & 1(12)\\
& & ad. LASSO & 0.4868(10) & 0.8228(12) & 0.9484(12) & 0.9848(12) && 0.5480(12) & 0.8756(12) & 0.9756(11) & 0.9950(12)\\ 
& & ISIS-SCAD & 0.5698(21) & 0.9468(37) & 0.9962(52) & 0.9998(66) && 0.6520(21) & 0.9804(37) & 0.9996(52) & 1(66)\\ 
\\
\multirow{4}{*}{$\Sigma_2$} & \multirow{4}{*}{0.5}  & Basic \alg & 0.8292(10) & 0.9710(11) & 0.9956(12) & 0.9992(12) &\multirow{4}{*}{0.55}& 0.8722(11) & 0.9830(11) & 0.9980(12) & 0.9998(12)\\   
& & LASSO & 0.4996(10) & 0.6410(11) & 0.7320(12) & 0.7884(12) && 0.5446(11) & 0.6860(11) & 0.7730(12) & 0.8302(12)\\
& & ad. LASSO & 0.5336(10) & 0.6870(11) & 0.7866(12) & 0.8408(12) && 0.5748(11) & 0.7352(11) & 0.8212(12) & 0.8730(12)\\ 
& & ISIS-SCAD & 0.5148(21) & 0.6552(37) & 0.7298(52) & 0.7726(66) && 0.5286(21) & 0.6770(37) & 0.7604(52) & 0.8040(66)\\ 
\\
& \multirow{4}{*}{0.5}  & Basic \alg & 0.9414(11) & 0.9998(11) & 1(11) & 1(11) &\multirow{4}{*}{0.55}& 0.9720(11) & 1(11) & 1(11) & 1(11)\\   
$\Sigma_3$ && LASSO & 0.5698(11) & 0.7392(11) & 0.8278(11) & 0.8710(11) && 0.6166(11) & 0.7844(11) & 0.8624(11) & 0.9036(11)\\
$\rho_1$=0.5 & &ad. LASSO & 0.6068(11) & 0.7862(11) & 0.8702(11) & 0.9100(11) && 0.6592(11) & 0.8266(11) & 0.8998(11) & 0.9340(11)\\ 
& & ISIS-SCAD & 0.6300(21) & 0.8084(37) & 0.8642(52) & 0.8966(66) && 0.6650(21) & 0.8292(37) & 0.8806(52) & 0.9094(66)\\ 
\\
&\multirow{4}{*}{0.5}  & Basic \alg & 0.6772(8) & 0.9618(10.5) & 0.9990(11) & 0.9998(11) &\multirow{4}{*}{0.55}& 0.7590(9) & 0.9816(11) & 0.9998(11) & 1(11)\\   
$\Sigma_3$ && LASSO & 0.5568(10) & 0.7602(11) & 0.8556(11) & 0.9028(11) && 0.6124(10) & 0.8074(11) & 0.8922(11) & 0.9294(11)\\
$\rho_1$=0.3 & &ad. LASSO & 0.5802(10) & 0.7980(10.5) & 0.8942(11) & 0.9328(11) && 0.6386(10) & 0.8432(11) & 0.9242(11) & 0.9532(11)\\ 
& & ISIS-SCAD & 0.6968(21) & 0.8916(37) & 0.9330(52) & 0.9558(66) && 0.7434(21) & 0.9124(37) & 0.9512(52) & 0.9702(66)\\
  \hline  \hline
\end{tabular}
\caption{Average accuracy with the median number of selected predictors in parentheses. The number of selected predictors is pre-determined for ISIS-SCAD.}
\label{simu1}
\end{center}
\end{table}
\end{landscape}

\subsection{The case of large $p$}
In this subsection,  the case  of large $p$  is considered and Two-stage Algorithm is applied for predictor screening.  For comparison purpose, we also present results using~Basic Algorithm and ISIS-SCAD.  The data generating   processes are  the same as in Subsection~\ref{sec:small_p_sim} but with different parameter   values. The covariance matrix $\Sigma$ can be $I_{p\times p}$, $\Sigma_2$ or $\Sigma_3$. 
We set $r^*=0.8$ for $\Sigma = I_{p\times p}$ and $r^*=0.3$ for $\Sigma = \Sigma_2$. When $\Sigma = \Sigma_3$, we consider $\rho_1 \in \{ 0.3, 0.5 \}$ and set $r^*=0.4$ for  $\rho_1=0.3$ and $r^*=0.3$ for  $\rho_1=0.5$. The sample size $n$ is 200. Several values of $p$ are considered: 
$p=34000P_0$, where   $P_0 \in \{ 1, 2, 4, 6, 8\}$. That is, $p=
34000, ~68000, ~136000, ~204000, ~272000$. Note that here we have chosen smaller $r^*$ values than those in the previous simulation study. Under such a setup, we can observe clearly that the screening accuracy decreases as $p$ increases for Basic Algorithm   and ISIS-SCAD, and we can further investigate whether the same phenomenon occurs for Two-stage Algorithm.

For each $\Sigma$ and each  $P_0 \in \{ 1, 2, 4, 6, 8\}$,  we carry out  ISIS-SCAD, Basic Algorithm and Two-stage Algorithm with $T \in \{ 5, 10, 15, 20 \}$  for 100 simulated data sets and compute the corresponding accuracy measures defined in~(\ref{acc_measure}). The average accuracy and the median number of selected predictors of each algorithm are given in Table~\ref{simu_psis}.  As shown in  Table~\ref{simu_psis},  for ISIS-SCAD  and  Basic Algorithm, the average screening accuracy  tends to decrease as $P_0$ increases.   For Two-stage Algorithm with $\Sigma=I_{p\times p}$ or $\Sigma_2$,  the average screening accuracy still decreases  as $P_0$ increases, but  not  as much as for ISIS-SCAD  and  Basic Algorithm. For Two-stage Algorithm with $\Sigma=\Sigma_3$, the screening accuracy does not always decrease as $P_0$ increases.  The average accuracy of Two-stage Algorithm is better than that of ISIS-SCAD and Basic Algorithm. Moreover, the screening accuracy of Two-stage Algorithm with $T=20$ and $P_0 \in \{ 2, 4, 6, 8\}$  is comparable to that of Basic Algorithm with $P_0=1$. Note that $P_0=1$ is a case falling into the category of $p<n^{1.97}$, where Basic Algorithm is theoretically justified by Theorem~\ref{thm:1} and Theorem~\ref{fa:consist}. Therefore, we conclude that the performance of Two-stage Algorithm is quite good for large $p$. The effect of $T$, the number of repeated random partitions, is also of  interest.  As the results in Table~\ref{simu_psis} indicate, the screening accuracy can be improved by increasing $T$. However, the improvement  becomes small when $T \geq 15$.
\begin{table}[ht] \renewcommand{\arraystretch}{1.1} \small
\addtolength{\tabcolsep}{-5.5pt} \tabcolsep=5pt
\begin{changemargin}{-1cm}{-1cm}
\begin{center}
    \begin{tabular}{cccccccc}
    \hline  \hline
$\Sigma$ & $r^*$ & & $p=34000$ & $p=68000$ & $p=136000$ & $p=204000$ & $p=272000$ \\
    \cline{4-8}
   \multirow{6}{*}{$I_{p \times p}$} &\multirow{6}{*}{0.8}&ISIS-SCAD   & 0.800(37) & 0.727(37) & 0.677(37) & 0.632(37) & 0.609(37)\\
   & & Basic \alg & 0.893(11) & 0.829(10) & 0.802(10) & 0.758(10) & 0.740(10)\\
   & & Two-stage \alg (T=5) & & 0.874(14)   & 0.869(18) & 0.837(21) & 0.796(23)\\
   & & Two-stage \alg (T=10) & & 0.902(17)   & 0.907(27) & 0.871(32.5) & 0.851(38)\\
   & & Two-stage \alg (T=15) & & 0.914(20)   & 0.917(32) & 0.884(41) & 0.873(47)\\
   & & Two-stage \alg (T=20) & & 0.922(22)   & 0.924(36) & 0.893(45.5) & 0.886(52.5)\\
   \\

     \multirow{6}{*}{$\Sigma_2$} &\multirow{6}{*}{0.3}&ISIS-SCAD & 0.558(37) & 0.587(37) & 0.537(37) & 0.541(37) & 0.518(37)\\
    && Basic \alg & 0.817(9) & 0.816(9)  & 0.757(8)    & 0.733(8.5) & 0.705(8)\\
    &&Two-stage \alg (T=5) & & 0.844(10.5) & 0.821(13) & 0.814(18)   & 0.807(20)\\
    &&Two-stage \alg (T=10) & & 0.844(11) & 0.821(17) & 0.814(24)   & 0.807(30.5)\\
    &&Two-stage \alg (T=15) & & 0.844(12) & 0.821(20) & 0.814(29)   & 0.807(39)\\
    &&Two-stage \alg (T=20) & & 0.844(13) & 0.821(22) & 0.814(34.5)   & 0.807(45)\\
   \\
      &\multirow{6}{*}{0.3}&ISIS-SCAD & 0.705(37) & 0.716(37) & 0.688(37) & 0.669(37) & 0.654(37)\\
    && Basic \alg & 0.911(10) & 0.918(10)  & 0.876(10) & 0.856(10) & 0.862(10)\\
   $\Sigma_3$ &&Two-stage \alg (T=5) & & 0.943(12)  & 0.928(14) & 0.930(18) & 0.946(20)\\
  $\rho_1=0.5$  && Two-stage \alg (T=10) & & 0.943(13)  & 0.928(17.5) & 0.930(24) & 0.946(30)\\
    && Two-stage \alg (T=15) & & 0.943(13.5)  & 0.928(20) & 0.930(30) & 0.946(37.5)\\
    && Two-stage \alg (T=20) & & 0.943(14)  & 0.928(23) & 0.930(34) & 0.946(42.5)\\
   \\
    &\multirow{6}{*}{0.4}&ISIS-SCAD & 0.848(37) & 0.831(37) & 0.810(37) & 0.797(37) & 0.792(37)\\
     && Basic \alg & 0.855(10) & 0.842(10)  & 0.792(9)  & 0.761(9) & 0.765(9)\\
   $\Sigma_3$  &&Two-stage \alg (T=5) & & 0.880(12)  & 0.877(13) & 0.865(16) & 0.877(19)\\
   $\rho_1=0.3$  &&Two-stage \alg (T=10) & & 0.880(13)  & 0.877(17) & 0.865(23) & 0.877(29)\\
     &&Two-stage \alg (T=15) & & 0.880(14)  & 0.877(20) & 0.865(28) & 0.877(36)\\
     &&Two-stage \alg (T=20) & & 0.880(15)  & 0.877(23) & 0.865(33) & 0.877(41)\\
   \hline  \hline
\end{tabular}
\caption{Average accuracy for the large $p$ case, with the median number of selected predictors in parentheses}
\label{simu_psis}
\end{center}
\end{changemargin}
\end{table}

\subsection{Heavy-tailed distribution and skewed distribution} \label{sec:heavy}
In previous simulation studies, we assume that all predictors are from a normal distribution. To check the performance of the proposed method  
when the predictor distribution deviates from normal,  we carry out more simulation experiments under a heavy-tailed distribution and a skewed distribution. 

We first consider the case where the predictors are from the $t$-distribution with four degrees of freedom, which is a heavy-tailed distribution. The predictors are independent in this case. The performance results of Basic Algorithm, LASSO and adaptive LASSO are presented in Table~\ref{simu_add1}. We find    that  the performance of Basic Algorithm is better than those of LASSO and adaptive LASSO  in such a case. 

Next,  we consider the case where  the predictors are from the skew normal distribution with location parameter 1, scale parameter 1.5 and slant parameter -8 (\cite{azz2013}). We generate the predictors independently using   the R package \verb+sn+. The simulation results are   given  in Table~\ref{simu_add1}. We also find  that the performance of Basic Algorithm is superior to  those of LASSO and adaptive LASSO, especially when the sample size is small. 
  \begin{table}[ht] \renewcommand{\arraystretch}{1.2} \small
\addtolength{\tabcolsep}{-5.5pt} \tabcolsep=6pt
\begin{center}
    \begin{tabular}{ccccccc}
    \hline  \hline
  Distribution & $ r^*$ &  & $n = 100$ & $n = 200$ & $n = 300$ & $n = 400$ \\
   \hline
   \multirow{6}{*}{$t$} & \multirow{3}{*}{0.91} & Basic \alg & 0.6044(9) & 0.9986(12) & 0.9998(12) & 1(12) \\
   && LASSO & 0.5202(10) & 0.8816(12) & 0.9876(12) & 0.9988(12)\\
   && ad. LASSO & 0.4494(10) & 0.8036(12) & 0.9430(12) & 0.9822(12)\\
   
   &\multirow{3}{*}{0.95} & Basic \alg & 0.7726(12) & 1(12) & 1(12) & 1(12) \\
   && LASSO & 0.5902(12) & 0.9462(12) & 0.9986(12) & 1(12)\\
   && ad. LASSO & 0.5082(12) & 0.8536(12) & 0.9720(12) & 0.9942(12)\\
   \\
   \multirow{6}{*}{skew normal} &\multirow{3}{*}{0.91} & Basic \alg & 0.5868(9) & 0.9988(12) & 0.9998(12) & 1(11) \\
   && LASSO & 0.4400(10) & 0.7792(12) & 0.9468(12) & 0.9920(11)\\
   && ad. LASSO & 0.4974(10) & 0.8254(12) & 0.9432(12) & 0.9752(11)\\
   &\multirow{3}{*}{0.95} & Basic \alg & 0.7726(12) & 1(12) & 1(11.5) & 1(11) \\
   && LASSO & 0.4996(12) & 0.8316(12) & 0.9772(11.5) & 0.9994(11)\\
   && ad. LASSO & 0.5516(12) & 0.8712(12) & 0.9710(11.5) & 0.9900(11)\\
   \hline  \hline
\end{tabular}
\caption{Average accuracy with the median number of selected predictors in parentheses when predictors are not normally distributed}
\label{simu_add1}
\end{center}
\end{table}

\subsection{Different threshold values}
In this subsection, we perform simulation experiments on Basic Algorithm to check   its performance under  different $\alpha$ values.  The  simulation settings in these experiments are the same as those in Subsection~\ref{sec:small_p_sim} unless otherwise stated.

In the previous simulation studies, we always set  $\alpha=0.5$ for the proposed methods. In order to check whether the $\alpha$ value significantly affects the screening performance, we  carry out the same experiment as in Subsection~\ref{sec:small_p_sim} with $\Sigma=I_{p\times p}$ using Basic Algorithm, but here with $\alpha = $0.2, 0.35, 0.5, 0.65 and 0.8.  
 The screening results are presented in Table~\ref{simu_add3}. We  find that  when $\alpha$ is larger,   the proposed screening method  can detect more of the important predictors, yet it includes more of the unimportant predictors as well.  
  When the sample size is small,  using a small $\alpha$ value in Basic Algorithm often leads to failure in detecting  some important predictors.  Therefore,  we suggest  setting $\alpha=0.5$, so that when the sample size is small,  Basic Algorithm can pick out reasonably many unimportant predictors   while keeping the important ones.

\begin{table}[ht] \renewcommand{\arraystretch}{1.2} \small
\addtolength{\tabcolsep}{-5.5pt} \tabcolsep=6pt
\begin{center}
    \begin{tabular}{cccccc}
    \hline  \hline
   $ r^*$ & $\alpha$ & $n = 100$ & $n = 200$ & $n = 300$ & $n = 400$ \\
   \hline
   \multirow{5}{*}{0.91} & 0.8 & 0.6392(14) & 0.9992(15) & 1(14) & 1(15) \\
   & 0.65 & 0.6242(11) & 0.9990(13) & 1(13) & 1(13)\\
   & 0.5 & 0.6112(9) & 0.9984(12) & 1(12) & 1(12)\\
   & 0.35 & 0.5468(7) & 0.9990(11) & 1(11) & 1(11)\\
   & 0.2 & 0.4482(4) & 0.9960(10) & 1(10) & 1(10)\\
    \hline
   \multirow{5}{*}{0.95} & 0.8 & 0.8036(17) & 1(15) & 1(14) & 1(15) \\
   & 0.65 & 0.7992(14) & 1(13) & 1(13) & 1(13)\\
   & 0.5 & 0.7768(12) & 1(12) & 1(11) & 1(12)\\
   & 0.35 & 0.7346(11) & 1(11) & 1(11) & 1(11)\\
   & 0.2 & 0.6436(10) & 1(10) & 1(10) & 1(10)\\
   \hline  \hline
\end{tabular}
\caption{Average accuracy with the median number of selected predictors in parentheses for three different $\alpha$ values when $\Sigma=I_{p\times p}$}
\label{simu_add3}
\end{center}
\end{table}
\section{Real data applications} \label{sec:real_data}
\subsection{The EEG data}
In this subsection, we apply our screening method to the EEG data set, available at \verb-https://archive.ics.uci.edu/ml/datasets/eeg+database-.
The original EEG data set was provided by Henri Begleiter at the Neurodynamics Laboratory at the State University of New York Health Center at Brooklyn. The original EEG data set contains experimental EEG data from 122 subjects. 77 of the 122 subjects were in the alcoholic group and 45 were in the control group. 
One of the objectives for analyzing this data set is to find predictors for classification.  
The data for each subject are from 30-120 trials, where in each trial for a subject,  measurements were taken from 64 electrodes at 256 time points. By taking the average measures over the trials, we obtain a data set that contains 256 $\times$ 64 average measurements for each of the 122 subjects. If in a trial for a subject the measurements from an electrode are all zeros, the data from that trial are excluded when computing the average measurements for the subject. Similarly, if the measurements from an electrode are all zeros for all trials for a subject, then the 256 average measurements corresponding to that electrode are set to zeros.
Next, we transform the 256 average measurements from each electrode for each subject to spline coefficients by fitting a cubic spline curve to the 256 time-point measurements using least squares and then extracting the coefficients. The B-spline basis functions are of degree  one with three interior knots at $1/4$, $2/4$, $3/4$  and boundary knots at 0 and 1. Using this transformation, we now have predictors of size $65\times 5=320$.

To evaluate the classification results, we use the Monte Carlo leave-$k$-out approach with $k=3$. The subjects are split into a training group of size 119 and a testing group of size 3 each time, where the training group consists of 75 subjects from the alcoholic group and 44 subjects from the control group. The random split is repeated 100 times.
For each split, we build a model for prediction based on the training data set and evaluate the prediction accuracy on the testing subject. For variable screening, we first adopt Basic Algorithm using bootstrap approximation with 500 bootstrap samples. Based on the selected predictors from Basic Algorithm, we then perform logistic regression with  lasso penalty and  select the predictors with nonzero coefficients. Lastly, we  build a classifier using logistic regression based on the predictors in the final selection. Here the logistic regression with  lasso penalty is carried out using the +cv.glmnet+ command in the R package \verb+glmnet+. We note that after applying the Basic Algorithm to each training data, the number of selected predictors ranges from 94 to 118, so  a regularized logistic regression is necessary to prevent overfitting. The average classification accuracy rate on test data over the 100 splits is 0.8233. 

\subsection{The 1000 Genomes Project data}
We analyze the 1000 Genomes Project phase 1 data~\cite{1AJR12} by examining the PCA plot and the prediction accuracy based on LDA modelling. This dataset collects the DNA variations from 1092 individuals from 14 populations in 4 continents: Africa, America, Asia and Europe. For each individual, 
 the variation measures of 36,781,560 loci on 22 chromosomes are provided.  Thus the data matrix is of size 1092 $\times$ 36,781,560.
 
\begin{figure}[ht]
\begin{center}
\includegraphics[width=16cm]{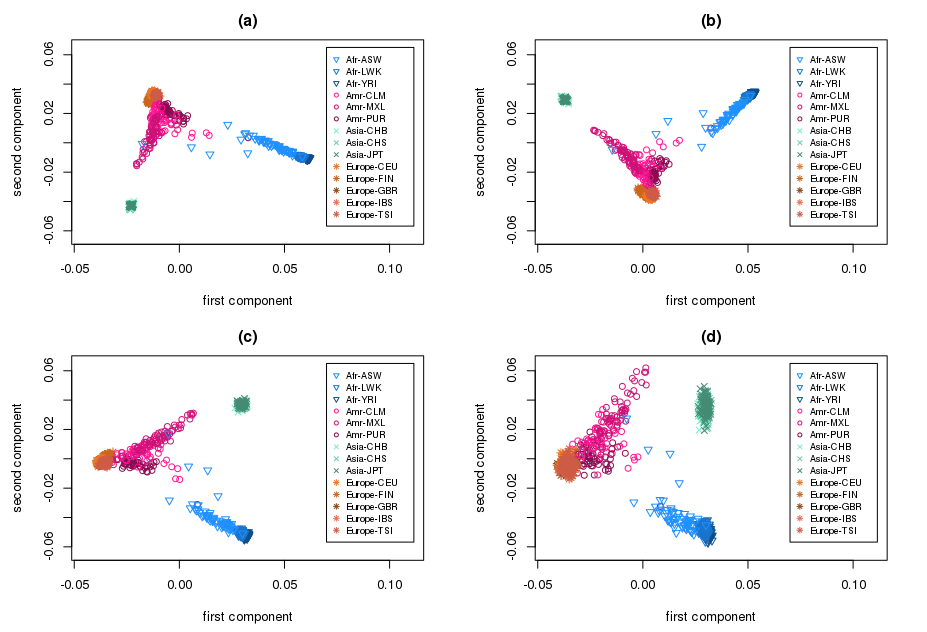}
\caption{The PCA plots for full data and for reduced data obtained by screening}
\label{svdres}
\end{center}
\end{figure}

In~\cite{nov2008}, it is demonstrated that the plot of the first two principal components of the data matrix reflects the geometric locations of individual origins. Here we include the plot in
panel (a) of Figure~\ref{svdres}.  We also perform 
variable screening on the 36,781,560 predictors using the indicator, whether the individual belongs to the $i$-th population, as the response variable for $i=1,\dots,14$. It leads to 14 subsets of selected predictors. We consider predictors that appear in at least $k$ times in these 14 subsets and let $A_k$ be the reduced data matrix for those selected predictors. 
The plots of the first two principal components using reduced data matrix $A_k$ for $k=5,7,9$, are shown in (b), (c), (d) of Figure~\ref{svdres}.   $A_5$, $A_7$ and $A_9$ have, respectively, 767579, 136665, and 20260  predictors.
We find that the PCA plots using the full data matrix and the reduced data matrices $A_5$, $A_7$, $A_9$ have  similar pattern.  We note that for this data set, our screening procedure stops at the first iteration for each of the 14 populations because the variables are highly correlated and there are enough variables selected  at the first iteration for stopping the screening procedure.  For each of the 14 populations, the number of selected variables ranges from 233,483 to 6,416,283. In general, when a predictor is selected by Basic Algorithm, predictors that are  highly correlated to the selected predictor are likely to be selected together.

Next, we split the data set into a training data set containing 983 individuals and a testing data set containing the remaining 109 individuals. The proportions of the individual populations for the training data are approximately the same as the proportions of the individual populations for the complete data.
The split is done twice. For each split, we first perform our screening process  with an additional forward selection step  based on the training data set for LDA model building  and evaluate  the predictive ability of the model based on the testing data set. We note that after performing our screening process, the numbers of selected predictors are 12,983,880 and 12,990,441 for the two splits, respectively. To prevent overfitting, we include the forward selection step.  After performing forward selection, the numbers of selected predictors are 183 and 179 for the two splits, respectively.
For the screening procedure with forward selection, we use whether the individual belongs to the $i$-th population as the response variable 
to obtain $A^*_i$: a set of selected predictors for $i=1$, $\ldots$, 14.  Then the predictors used for LDA modelling are the predictors in $\cup_{i=1}^{14} A^*_i$.
The prediction accuracy  rates for the two splits are 0.4587 and 0.5505 respectively.  The low accuracy rates are probably due to the fact that some of the 14 populations are close to each other and can be grouped together.  Based on the PCA plots,  we put the 14 populations into the European group, the African group, the American group and the Asian group, and build the 4-group LDA model based on the training data. The resulting prediction accuracy rates on the testing data are 0.8991 and 
0.8716 respectively for the two random splits.  





\section{Concluding remarks} \label{sec:conclusion}
 In this article, we have considered   the problem of predictor screening  for linear regression in high dimensions.  We propose a distribution-based thresholding rule and give two theorems to justify the use of the proposed thresholding rule for moderate sized $p$. Theorem 1 states that screening based on the proposed thresholding rule will include important predictors with probability tending to one, and Theorem 2 states that the probability that too many unimportant predictors being selected can be controlled. Based on the proposed thresholding rule, we prescribe the Basic Algorithm for iterative variable screening. As the two main theorems are no longer valid for large $p$, we further propose a Two-stage Algorithm for large $p$. It includes a partitioning step that partitions predictors into smaller subgroups so that the Basic Algorithm can be applied to each subgroup. Next, results from repeated random partitions are collected and integrated to form the final estimate of active predictor set.  Simulation studies show that our method works well and outperforms some renowned competitors. Real data applications are also presented.

\section{Proofs} \label{sec:proof}
In this section, we give proofs for Theorem~\ref{thm:1}, Corollary~\ref{cor:1} and Theorem~\ref{fa:consist}.
\subsection{Proof of Theorem 1} 

Recall that  $Y_n =(\y_1, \ldots, \y_n)^\top$ is the observation vector of the response variable $\y$.
Suppose that given $Y_n$, $\X^*_1$, $\ldots$, $\X^*_p$ are independent, where
$\X^*_j = (\x^*_{1j}$, $\ldots$, $\x^*_{nj})^\top$ is a random sample from 
$\hat{F}_j$  for $j=1$, $\ldots$, $p$. Let  $\bar{\y} = \sum_{i=1}^n \y_i/n$ and let $\bar{\x}^*_j= \sum_{i=1}^n \x^*_{ij}/n$ for $j=1$, $\ldots$, $p$.
Let $\xmat$ denote the $n \times p$ data matrix whose $(i,j)$-th component is $\x_{ij}$. Then
\begin{eqnarray*}
 && G(t|Y_n, \hat{F}_1, \ldots, \hat{F}_p) \\
  && =P \left(  \left. \max_{ 1 \leq j \leq p } \left| \frac{ \sum_{i=1}^n (\x^*_{ij} - \bar{\x}^*_j)(\y_i - \bar{\y}) }{
   \left(  \sum_{i=1}^n (\x^*_{ij} - \bar{\x}^*_j)^2 \right)^{1/2}  \left( \sum_{i=1}^n (\y_i - \bar{\y})^2 \right)^{1/2}
  } \right| \leq t \right| Y_n, \xmat \right) \\
  && = \prod_{j=1}^p P \left( \left.  \left| \frac{ \sum_{i=1}^n (\x^*_{ij} - \mu_{j,n} )(\y_i - \bar{\y}) }{
   \left(  \sum_{i=1}^n (\x^*_{ij} - \bar{\x}^*_j)^2 \right)^{1/2}  \left( \sum_{i=1}^n (\y_i - \bar{\y})^2 \right)^{1/2}
  } \right| \leq t \right| Y_n, \xmat  \right), \\
\end{eqnarray*}
where  for  $j=1$, $\ldots$, $p$,  $\mu_{j,n} = \sum_{i=1}^n \x_{ij}/n$, and
\begin{eqnarray*}
&& P \left( \left.  \left| \frac{ \sum_{i=1}^n (\x^*_{ij} - \mu_{j,n})(\y_i - \bar{\y}) }{
   \left(  \sum_{i=1}^n (\x^*_{ij} - \bar{\x}^*_j)^2 \right)^{1/2}  \left( \sum_{i=1}^n (\y_i - \bar{\y})^2 \right)^{1/2}
  } \right| \leq t \right| Y_n, \xmat  \right) \\
&& \geq \underbrace{   P \left( \left.  \left| \frac{ \sum_{i=1}^n (\x^*_{ij} -\mu_{j,n})(\y_i - \bar{\y}) }{
   \sqrt{ n \sigma_{j,n}^2 }   \left( \sum_{i=1}^n (\y_i - \bar{\y})^2 \right)^{1/2}
  } \right| \leq t (1-\delta_1) \right| Y_n, \xmat  \right) }_{I_j} \\
&& \hspace{0.5cm} - \underbrace{  P \left(  \left. \left| \frac{ \sqrt{ n \sigma_{j,n}^2 }  }{
    \left(  \sum_{i=1}^n (\x^*_{ij} - \bar{\x}^*_j)^2 \right)^{1/2}} \right| > \frac{1}{1 - \delta_1}  \right| \xmat \right) }_{II_j}
\end{eqnarray*}
for $\delta_1 \in (0,1)$ and $\sigma_{j,n}^2 = \sum_{i=1}^n (\x_{ij}-\mu_{j,n})^2/n$.

Let $\Phi$ denote the CDF for $N(0,1)$. Below we will give bounds for $I_j$ and $II_j$ respectively by comparing them with probabilities related to $N(0,1)$. We will first show that $\max_{1 \leq j \leq p} II_j = O_p(1/n^2)$.
For $\delta_2 \in (0, 1-(1-\delta_1)^2)$, we have
\begin{eqnarray*}
&& II_j = P \left(   \left. \left| \frac{ \sqrt{ n \sigma_{j,n}^2 }  }{    \left(  \sum_{i=1}^n (\x^*_{ij} - \bar{\x}^*_j)^2 \right)^{1/2}} \right| > \frac{1}{1 - \delta_1}  \right| \xmat \right) \\
&& \leq P \left( \left. \frac{ \sum_{i=1}^n (\x^*_{ij}- \mu_{j,n})^2 }{n \sigma_{j,n}^2  } < (1-\delta_1)^2 - \delta_2 \right| \xmat \right)
 + P \left( \left. \frac{n (\bar{\x}^*_j- \mu_{j,n})^2 }{n \sigma_{j,n}^2  } > \delta_2 \right| \xmat \right). \\
 \end{eqnarray*}
 From Theorem  2 in Osipov and Petrov~\cite{osipov1967},  
 \begin{eqnarray*}
 &&P \left( \left. \frac{ \sum_{i=1}^n (\x^*_{ij}- \mu_{j,n})^2 }{n \sigma_{j,n}^2  } < (1-\delta_1)^2 - \delta_2 \right| \xmat \right)
 + P \left( \left. \frac{n (\bar{\x}^*_j- \mu_{j,n})^2 }{n \sigma_{j,n}^2  } > \delta_2 \right| \xmat \right) \\
 \end{eqnarray*}
 can be expressed as
 \begin{eqnarray*}
 &&  \Phi( \sqrt{n} (  (1-\delta_1)^2 - \delta_2 -1)/\sqrt{v_{j,n}}) + (1-2\Phi(\sqrt{n\delta_2})) + a_{n,j} S_{n,1}
\end{eqnarray*}
 for $1 \leq j \leq p$,  where 
 \[
 v_{j,n} = Var \left( \left. \frac{ (\x^*_{ij}-\mu_{j,n})^2}{ \sigma_{j,n}^2} \right| \xmat \right) = \frac{1}{n} \sum_{i=1}^n \left( \frac{ (\x_{ij}-\mu_{j,n})^2}{ \sigma_{j,n}^2} -1 \right)^2,
\]
$\max_{1\leq j \leq p} |a_{n,j}| =O(1/n^2)$,
\[
 S_{n,1} = \max_{1 \leq j \leq p} \frac{1}{n} \sum_{i=1}^n \left| \frac{(\x_{ij} - \mu_{j,n})^2 }{ \sigma_{j,n}^2 } - 1 \right|^3 
   + 
  \max_{1 \leq j \leq p} \frac{1}{n} \sum_{i=1}^n \left| \frac{\x_{ij} - \mu_{j,n}}{ \sigma_{j,n}}  \right|^3.
  \]
Since $\max_{1\leq j \leq p} v_{j,n} = O_p(1)$ and $S_{n,1} = O_p(1)$, we have $\max_{1 \leq j \leq p} II_j = O_p(1/n^2)$.

To control $I_j$, note that
\begin{eqnarray}
&& \left| P \left( \left.  \left| \frac{ \sum_{i=1}^n (\x^*_{ij} -\mu_{j,n})(\y_i - \bar{\y}) }{
   \sqrt{ n \sigma_{j,n}^2 }   \left( \sum_{i=1}^n (\y_i - \bar{\y})^2 \right)^{1/2}
  }  \right| \leq t (1-\delta_1) \right| Y_n, \xmat \right) 
   -  \left(1-2\Phi(   t (1-\delta_1)\sqrt{n}) \right) \right| \nn \\
   && \leq \left| P \left( \left.   \frac{ \sum_{i=1}^n (\x^*_{ij} -\mu_{j,n})(\y_i - \bar{\y}) }{
   \sqrt{ n \sigma_{j,n}^2 }   \left( \sum_{i=1}^n (\y_i - \bar{\y})^2 \right)^{1/2}
  }   \leq t (1-\delta_1) \right| Y_n, \xmat \right) - \left(1-2\Phi(   t (1-\delta_1)\sqrt{n}) \right)  \right| \nn \\
  && + \left| P \left( \left.   \frac{ -\sum_{i=1}^n (\x^*_{ij} -\mu_{j,n})(\y_i - \bar{\y}) }{
   \sqrt{ n \sigma_{j,n}^2 }   \left( \sum_{i=1}^n (\y_i - \bar{\y})^2 \right)^{1/2}
  }   \leq t (1-\delta_1) \right| Y_n, \xmat \right) -  \left(1-2\Phi(   t (1-\delta_1)\sqrt{n}) \right)  \right| \nn \\
&& \leq  \left( \frac{ 2C_1 (\sum_{i=1}^n |\x_{ij} - \mu_{j,n}|^3/n)\sum_{i=1}^n |\y_i - \bar{\y}|^3 }{ \sigma_{j,n}^3 \left( \sum_{i=1}^n |\y_i - \bar{\y}|^2   \right)^{3/2} } \right) \left( \frac{1}{ 1+ \sqrt{n}\cdot t (1-\delta_1)   }  \right)^3 \nn \\
 && \leq    \left( \frac{ 2C_1 S_{n,1} \sum_{i=1}^n |\y_i - \bar{\y}|^3 /n}{ \sqrt{n} \left( \sum_{i=1}^n |\y_i - \bar{\y}|^2/n   \right)^{3/2} } \right) \left( \frac{1}{ 1+ \sqrt{n}\cdot t (1-\delta_1)   }  \right)^3
\label{eq:boundI}
\end{eqnarray}
for some absolute constant $C_1$. Here the last inequality follows from Theorem 2 in~\cite{osipov1967}. Note that in (\ref{eq:boundI}),  the quantity 
\[
 \frac{S_{n,1} \sum_{i=1}^n |\y_i -\bar{\y}|^3 /n}{ \left(\sum_{i=1}^n (\y_i - \bar{\y})^2 /n \right)^{3/2} } = O_p(1),
\]
so
\[
 \max_{1 \leq j \leq p}  \left|  I_j
   -   \left(1-2\Phi(   t (1-\delta_1)\sqrt{n}) \right) \right| = O_p(1/n^2),
\]
which, together with the fact that $\max_{1 \leq j \leq p} II_j = O_p(1/n^2)$, implies that
\[
 G(t|Y_n, \hat{F}_1, \ldots, \hat{F}_p)  \geq \prod_{j=1}^p ( I_j - II_j).
\]
Therefore, for $t>0$, $G(t|Y_n, \hat{F}_1, \ldots, \hat{F}_p) \goes 1$ in probability as $n \goes \infty$, which implies that for $t>0$ and $\alpha \in (0,1)$,
\[
  \lim_{n\goes \infty} P(1-\alpha  < G(t/2|Y_n, \hat{F}_1, \ldots, \hat{F}_p))  = 1.
\]
Since
\[
P(t_{\alpha,n} < t) \geq P(t_{\alpha,n} \leq t/2 ) \geq  P(1-\alpha  < G(t/2|Y_n, \hat{F}_1, \ldots, \hat{F}_p)),
\]
we have $P(t_{\alpha,n} < t) \goes 1$ as $n \goes \infty$.
The proof of Theorem~\ref{thm:1} is completed.

\subsection{Proof of Corollary~\ref{cor:1}}

For $t>0$, let $A_n(t)$ denote the event that the  absolute sample correlations between $Y_n$ and all important predictors exceed $t$. From Theorem~\ref{thm:1}, 
 \[ 
 \lim_{n\goes \infty} P(t_{\alpha,n} < t )  = 1,
\]
 so when  $t_{\alpha,n}$ is used as the screening threshold,
 $\lim_{n\goes \infty} P( \{ \cA \not\subset {\hat \cA}_n \} \cap A_n(t) )= 0$. From the assumption that $\cA$ is fixed and finite, $\lim_{n\to\infty} P(A_n(t)) =1$ for some $t >0$, so
we have $\lim_{n\to\infty} P\left\{\cA  \subset   {\hat \cA}_n \right\} =1$.

When $z(n,p,\alpha)$ is used as the screening threshold, 
 under the condition that $p \leq Cn^{2-\delta}$ for some constant $C>0$ and some small $\delta >0$, it can be shown that $$\lim_{n\goes \infty} z(n, p, \alpha) =0.$$ Thus, 
 when  $z(n,p,\alpha)$ is used as the screening threshold,
 $\lim_{n\goes \infty} P( \{ \cA \not\subset {\hat \cA}_n \} \cap A_n(t) )= 0$. From the assumption that $\cA$ is fixed and finite, $\lim_{n\to\infty} P(A_n(t)) =1$ for some $t >0$, so
we have $\lim_{n\to\infty} P\left\{\cA  \subset   {\hat \cA}_n \right\} =1$.
The proof  of Corollary~\ref{cor:1} is completed.
 
\subsection{Proof of Theorem~\ref{fa:consist}}
Proof. For $j=1$, $\ldots$, $p-\K$, let $\I_j=1$ if the $j$-th unimportant predictor is selected and 0 otherwise. It is clear that $\I_1$, $\ldots$, $\I_{p-\K}$ are
conditionally independent given $\y_1$, $\ldots$, $\y_n$: the observations of the response variable. In addition, 
the conditional probability $P(\I_j=1|\y_1, \ldots, \y_n)$ is
\begin{eqnarray*}
&& P\left( \left| \frac{\sum_{i=1}^n (\y_i - \bar{y})( \w_i - \bar{\w}) }{ \sqrt{\sum_{i=1}^n (\y_i - \bar{y})^2} \sqrt{\sum_{i=1}^n ( \w_i - \bar{\w})^2} }  \right| > \frac{c(p)}{\sqrt{n}} |\y_1, \ldots, \y_n \right) \\
 && = P \left( \left| \frac{Z_1}{ \sqrt{\sum_{\ell=1}^{n} Z_{\ell}^2  - Z_2^2} } \right| > \frac{c(p)}{\sqrt{n}} |\y_1, \ldots, \y_n \right),
\end{eqnarray*}
where $\bar{\y} = \sum_{i=1}^n \y_i/n$,
$\w_1$, $\ldots$, $\w_n$ are IID observations of the $j^{\mbox{th}}$ unimportant predictor,  $\bar{\w} = \sum_{i=1}^n \w_i/n$, $c(p)$ is given in (\ref{eq:facp}),
\[
 Z_1 = \frac{\sum_{i=1}^n (\y_i - \bar{y}) \v_i}{ \sqrt{\sum_{i=1}^n (\y_i - \bar{y})^2} } 
\]
with $\v_i = (\w_i - E(\w_i))/\sqrt{\Var(\w_i)}$, $Z_2 = \sum_{i=1}^n \v_i/\sqrt{n}$,
and $Z_{3}$, $\ldots$, $Z_n$ are IID $N(0,1)$ variables that are conditionally independent of $(Z_1, Z_2)$ given $\y_1$, $\ldots$, $\y_n$. Since $Z_1$ and $Z_2$ are IID $N(0,1)$ given $\y_1$, $\ldots$, $\y_n$, we have $P(\I_j=1|\y_1, \ldots, \y_n)= p_1$, where $p_1$ is given in (\ref{eq:fap1}). Since $p_1$ does not depend on $\y_1$, $\ldots$, $\y_n$, the distribution for $\sum_{j=1}^{p-\K} \I_j$ is $Bin(p-\K, p_1)$.

Next, we will provide an upper bound for $p_1$. Note that
\begin{eqnarray*}
 p_1 & = & P \left( \left| \frac{Z}{ \sqrt{Z^2/n + U_n/n} } \right| >  c(p)  \right) \\
 & \leq & P( |Z|/\sqrt{1-\delta_0} > c(p) \mbox{ and } U_n/n \geq 1-\delta_0) + P(U_n/n < 1 - \delta_0) \\
 & \leq & 2 ( 1- \Phi(c(p)\sqrt{1-\delta_0} )) + P(U_n < n(1-\delta_0))
\end{eqnarray*}
for $\delta_0 \in (0,1)$, where $\Phi$ is the CDF of $N(0,1)$. Let $u_p = 0.5(1-(1-\alpha)^{1/p})$ and $\beta=1/8$. Take
\[
p_1^* = 2 ( 1- \Phi(c(p)\sqrt{1-u_p^{\beta}} )) + P(U_n < n(1-u_p^{\beta})),
\] then $p_1 \leq p_1^*$. We will show that $\lim_{n\goes \infty} pp_1^* = -\ln(1-\alpha)$ by proving (\ref{eq:fap12}) and (\ref{eq:fap13}):
\begin{equation} \label{eq:fap12}
 \lim_{n\goes \infty} 2p ( 1- \Phi(c(p)\sqrt{1-u_p^{\beta}} )) = -\ln(1-\alpha)
\end{equation}
and
\begin{equation} \label{eq:fap13}
 \lim_{n\goes \infty} pP(U_n < n(1-u_p^{\beta})) = 0.
\end{equation}

To prove (\ref{eq:fap13}), note that
\[
 P( n-2 - U_n > n u_p^{\beta} - 2 ) \leq \frac{Ee^{t (n-2 - U_n)}}{e^{t(n u_p^{\beta} - 2)}}  = \frac{e^{t(n-2)} (1+2t)^{-(n-2)/2}}{e^{t(n u_p^{\beta} - 2)}}
\]
 Take $t = 1/\sqrt{2(n-2)}$, then $\lim_{n\goes \infty} e^{t(n-2)} (1+2t)^{-(n-2)/2} = e^{1/2}$ and
\[
 t(n u_p^{\beta} - 2) = \frac{1}{\sqrt{2(n-2)}} \left( n \left( \frac{-\ln(1-\alpha)}{2p} \right)^{\beta} c_1(p) -2 \right),
\]
where 
\[
 c_1(p) = \frac{u_p^{\beta}}{ (-\ln(1-\alpha)/(2p) )^{\beta} } =  \left( \frac{0.5(1-(1-\alpha)^{1/p})}{-\ln(1-\alpha)/(2p)} \right)^{\beta} \goes 1
\]
as $n \goes \infty$, so $t(n u_p^{\beta} - 2) > c^* n^{1/4}$ for some constant $c^*>0$, which gives
\[
 \lim_{n\goes \infty} p  P( n-2 - U_n > n u_p^{\beta} - 2 ) \leq e^{1/2} \lim_{n\goes \infty} p e^{-t c^* n^{1/4}} =0
\]
and we have (\ref{eq:fap13}).
 
To prove (\ref{eq:fap12}), note that
\begin{eqnarray*}
 && \lim_{n\goes \infty} 2p ( 1- \Phi(c(p)\sqrt{1-u_p^{\beta}} )) \\
 &&= \lim_{n\goes \infty} \frac{ z^{\delta_1} (1- \Phi( z\sqrt{1-\delta_1} )) }{ u_p^{1-\delta_1} } \lim_{n\goes \infty} 2 pz^{-\delta_1} u_p^{1-\delta_1},
\end{eqnarray*}
where $\delta_1 = u_p^{\beta}$, $z=\Phi^{-1}(1-u_p)$ and the limit 
\begin{eqnarray*}
&& \lim_{n\goes \infty} \frac{ z^{\delta_1} (1- \Phi( z\sqrt{1-\delta_1} )) }{ u_p^{1-\delta_1} } \\
 &&= \lim_{n\goes \infty} \frac{ z^{\delta_1} (1- \Phi( z\sqrt{1-\delta_1} )) }{ (1-\Phi(z))^{1-\delta_1} } \\
 && = \lim_{n\goes \infty} \frac{z^{\delta_1} \phi( z\sqrt{1-\delta_1} )}{z\sqrt{1-\delta_1}} \left( \frac{z}{\phi(z)} \right)^{1-\delta_1} =1,
\end{eqnarray*}
where $\phi$ is the density for $N(0,1)$. Thus
\begin{eqnarray*}
 && \lim_{n\goes \infty} 2p ( 1- \Phi(c(p)\sqrt{1-u_p^{\beta}} )) \\
 && = \lim_{n\goes \infty} 2 p u_p \lim_{n\goes \infty} z^{-\delta_1} u_p^{-\delta_1} \\
 && =   \lim_{n\goes \infty} 2p \frac{-\ln(1-\alpha)}{2p} \frac{0.5(1-(1-\alpha)^{1/p})}{  -\ln(1-\alpha)/(2p)  } \lim_{n\goes \infty} z^{-\delta_1} u_p^{-\delta_1} \\
 && =-\ln(1-\alpha)\lim_{n\goes \infty} z^{-\delta_1} u_p^{-\delta_1} \\
 && = -\ln(1-\alpha) \lim_{n\goes \infty} \exp \left( - (1-\Phi(z))^{\beta} \ln( z (1-\Phi(z)) ) \right) \\
 && =   -\ln(1-\alpha) \lim_{n\goes \infty} \exp \left( - (1-\Phi(z))^{\beta} \ln(\phi(z) ) \right) = -\ln(1-\alpha) e^0 = -\ln(1-\alpha)
\end{eqnarray*}
and (\ref{eq:fap12}) holds.

From (\ref{eq:fap12}) and (\ref{eq:fap13}), $\lim_{n\goes \infty} p p_1^* = -\ln(1-\alpha)$. The rest results stated in Theorem~\ref{fa:consist} follow from the Poisson approximation for a Binomial distribution, and the fact that for a positive integer $n_0$ and for $r=0$, $\ldots$, $n_0$, the probability that a $Bin(n_0,p_0)$ variable is at least $r$ increases as $p_0$ increases.
The proof of Theorem~\ref{fa:consist} is completed.

\end{document}